\documentstyle[11pt,newpasp,twoside,epsf]{article}
\def\edcomment#1{\iffalse\marginpar{\raggedright\sl#1\/}\else\relax\fi}
\reversemarginpar

\begin{document}
\title{Detection of anisotropy in the 
Cosmic Microwave Background at horizon and sub-horizon scales 
with the BOOMERanG experiment}
 \author{P. de Bernardis$^{1}$,
P.A.R.Ade$^{2}$,  
J.J.Bock$^{3}$, 
J.R.Bond$^{4}$, 
J.Borrill$^{5,6}$, 
A.Boscaleri$^{7}$, 
K.Coble$^{8}$, 
B.P.Crill$^{9}$,
G. De Gasperis $^{10}$, 
G. De Troia$^{1}$, 
P.C.Farese$^{8}$, 
P.G.Ferreira$^{11}$, 
K.Ganga$^{9,11}$, 
M.Giacometti$^{1}$, 
E.Hivon$^{9}$, 
V.V.Hristov$^{9}$, 
A.Iacoangeli$^{1}$, 
A.H.Jaffe$^{6}$, 
A.E.Lange$^{9}$, 
L.Martinis$^{13}$, 
S.Masi$^{1}$, 
P.Mason$^{9}$, 
P.D.Mauskopf$^{14}$, 
A.Melchiorri$^{1}$, 
L.Miglio$^{1,15}$, 
T.Montroy$^{8}$, 
C.B.Netterfield$^{15}$, 
E.Pascale$^{7}$, 
F.Piacentini$^{1}$, 
D.Pogosyan$^{4}$, 
F.Pongetti$^{16}$,
S.Prunet$^{4}$, 
S.Rao$^{16}$, 
G.Romeo$^{16}$, 
J.E.Ruhl$^{8}$, 
F.Scaramuzzi$^{13}$, 
D.Sforna$^{1}$, 
N.Vittorio$^{10}$ 
 }


\affil{$^1$ Dipartimento di Fisica, Universit\'a di Roma La Sapienza, Roma, Italy;
$^2$ Dept. of Physics, Queen Mary and Westfield College, London, UK; $^3$ Jet Propulsion Laboratory, Pasadena, CA, USA; $^4$ CITA University of Toronto, Canada; 
$^5$ NERSC-LBNL, Berkeley, CA, USA; $^6$ Center for Particle Astrophysics, 
Univ. of California at Berkeley, USA; $^7$ IROE - CNR, Via Panciatichi 64, 50127 Firenze, Italy; $^8$  Department of Physics, Univ. of California at Santa Barbara, USA; 
$^9$ California Institute of Technology, Pasadena, USA; $^{10}$ Dipartimento di Fisica, Universit\'a di Roma Tor Vergata, Roma, Italy; $^{11}$ Astrophysics, University 
of Oxford, UK; $^{12}$ PCC, College de France, Paris, France; $^{13}$ ENEA Centro 
Ricerche di Frascati, Italy ; $^{14}$  Physics and Astronomy Dept, Cardiff University, UK; $^{15}$  Departments of Physics and Astronomy, Univ. of Toronto, Canada; $^{16}$ 
Istituto Nazionale di Geofisica, Roma, Italy}

\begin{abstract}
BOOMERanG has recently resolved  
structures on the last scattering surface at redshift $\sim$ 1100
with high signal to noise ratio.
We review the technical advances which
made this possible, and we focus on the current results 
for maps and power spectra, with special attention to the
determination of the total mass-energy density in the
Universe and of other cosmological parameters.
\end{abstract}

\section{Introduction}

A wide class of cosmological models predicts a harmonic series 
of peaks in the power spectrum of the Cosmic Microwave Background (CMB). 
These are the imprint of the acoustic oscillations inside the 
horizon in the primeval plasma. 
At recombination ($z \sim 1100$, $t \sim 300000 h^{-1}$ years) 
the acoustic horizon subtends an angle of roughly 
$1^o$, corresponding to 
a first peak at multipoles $\ell \sim 200$ in the spherical
harmonic expansion of the CMB temperature. 
A high confidence measurement of this peak came from
analysis of the high quality image of about 3$\%$ of the sky 
obtained in the long duration flight of BOOMERanG (de Bernardis et al. 2000; 
Lange et al. 2000). An important confirmation arrived soon after from 
the MAXIMA experiment (Hanany et al. 2000; Balbi et al. 2000).
Here we focus on the BOOMERanG experiment, which has produced 
a wide ($\sim 1800$ square degrees), faithful image of the CMB at angular scales smaller 
than 10$^o$. 

\section{The Instrument}

BOOMERanG is a scanning telescope, featuring three important improvements
with respect to previous experiments. First, BOOMERanG uses a
long duration (7 to 14 days) stratospheric ($\sim 38$ km) 
balloon flight around Antarctica.
Long integrations on a wide sky region are obtained, along with 
careful and extensive tests for systematic effects.
In addition, flying during the austral summer from Antarctica,
the lowest Galactic contamination region of the sky (Schlegel et al. 1998) 
is visible at an azimuth almost perfectly opposite to the azimuth of the Sun. 
This simplifies the necessary shielding required for thermal and
optical reasons. Second, BOOMERanG uses a very sensitive
total power receiver, based on spider-web bolometers (Mauskopf et al. 1997)
cooled to 0.28 K with a custom cryogenic system (Masi et al. 1998; 1999).
The power detected from one direction is compared to the power from
contiguous directions by slowly scanning the telescope (1$^o/s$ to 2$^o/s$
in azimuth). This strategy is enabled by the intrinsic stability of the
readout electronics (a low noise AC bridge) and of the detectors. The full
payload is gently moved, avoiding mechanical choppers 
and the related inefficiencies and slowly varying offsets. Third, 
the focal plane is multiband, with 8 pixels and
4 colors (90, 150, 240, 410 GHz) strategically located with respect
to the scan direction in order to have several 
temporal and spectral confirmations of the detected structures. 
We track the azimuth of the best sky region while scanning at constant elevation, 
thus obtaining higly crosslinked maps (see fig.1).
\begin{figure}
\plotone{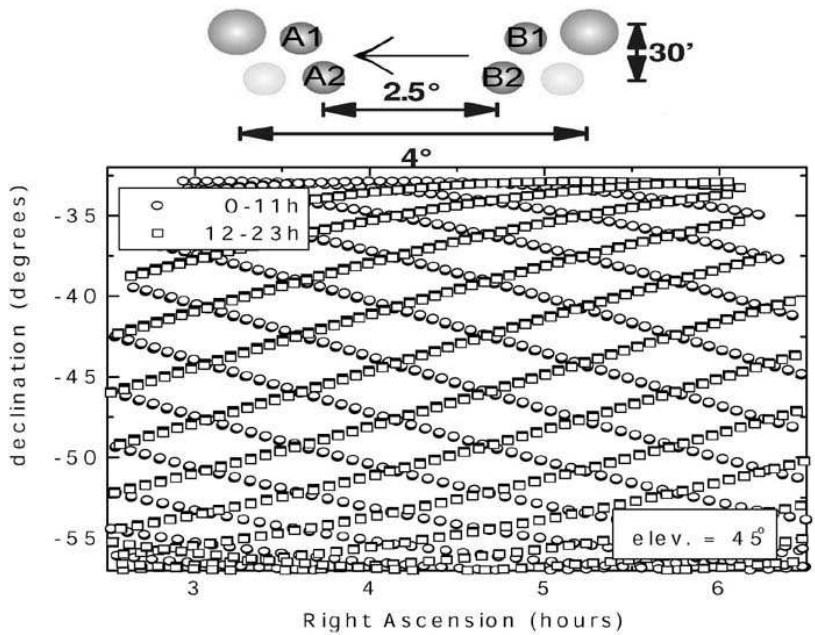}
\caption{
Beams projected in the sky from different detectors in the
focal plane of BOOMERanG, and scan strategy. The two larger 
beams are for the single-mode 90GHz detectors,
while the two detectors immediately below are single-mode 150GHz
detectors. The beams A1..B2 are for the four multiband photometers,
each observing simultanously at 150, 240 and 410 GHz.
Slow azimuth scans ($\pm 30^o$, 1$^o/s$ or 2$^o/s$)
are continuosly performed at constant elevation, while the center of the
scan tracks the azimuth of the lowest foreground region.
Only one scan every hour is shown in the lowest part of the 
figure. A structure detected in the forward scan in 
A1 will be detected a few seconds later in B1, and 30 seconds
later in time reverse in B1 and A1 during the return scan.
Due to sky rotation, the same sequence of events will be detected
a few minutes later in A2 and B2. Hours later, the same
sky pixel is observed again with a different inclination
of the scan path. All this is repeated every day for 10 days.
}
\end{figure}
The main features of the detectors array (as measured in flight)
are listed in table 1 (Crill et al. 2000, Piacentini et al. 2000)
\section{The Data}
The instrument was flown from McMurdo on Dec.29, 1998, and remained at an altitude
higher than 37 km for 10.6 days, circumnavigating Antarctica along
the 78$^o$S parallel and performing nominally. 
We devoted 106 hours of the flight to scans in the
best region at $1^o/s$, 82 hours to scans at $2^o/s$; the rest of the time 
to calibration, observation of selected sources and diagnostics. Data were
edited for instrumental events (calibration lamp flashes, telemetry glitches,
elevation changes, bias changes, bias trimming, cosmic rays events). About
5$\%$ of the data in each of the channels was flagged in this process, 
replaced with a constrained realization of noise, and 
not used for further analysis. 
\begin{table}
\caption{Important characteristics of BOOMERanG}
\begin{tabular}{lccccc}
\tableline
Parameter                      	    & 90GHz    & 150GHz     & 240GHz     & 410GHz     & \\
\tableline
number of detectors                     & 2        & 6          & 4          & 4          & \\
best $NET (\mu K_{CMB} s^{1 \over 2})$  & 140      & 130        & 170        & -          & \\
typical FWHM (arcmin)                   & 19       & 10         & 14         & 13         & \\
\tableline
\tableline
\end{tabular}
\end{table}
A preliminary pointing solution has been 
obtained from the data of the sun sensors, laser gyroscopes and differential
GPS. The residual jitter is of the order or 3' rms as checked on the
apparent location of known Galactic sources. The responsivity calibration
was obtained from the signal of the CMB dipole, visible in the data as a
scan synchronous triangular signal. 
To check for systematic effects at the scan
frequency, we calibrate separately the data at $1^o/s$ and the data
at $2^o/s$. The spread of the results is of the order of 10$\%$, due
to scan sysnchronous noise, which is stronger at $2^o/s$ but is
efficiently monitored using the 410 GHz channel data. So we attach to
the responsivity calibration a conservative error of 10$\%$, dominated by systematic
effects. The beam was measured before the flight with a tethered 
thermal source in the far field of the telescope. A beam model was
constructed from these observations, and checked in flight on compact
HII regions like RCW38 (see table 1). 
The preliminary uncertainty on the beam FWHM is of the order 
of 10$\%$. This hardly affects the power spectrum measurements at $\ell < 600$ 
(see de Bernardis et al. 2000).
In order to reduce the effect of drifts and 1/f noise, we 
high-pass filter the data in the time domain before making the map.
This effectively corresponds to removing all the structures larger than
10$^o$ along the scan direction. 
\begin{figure}
\plotone{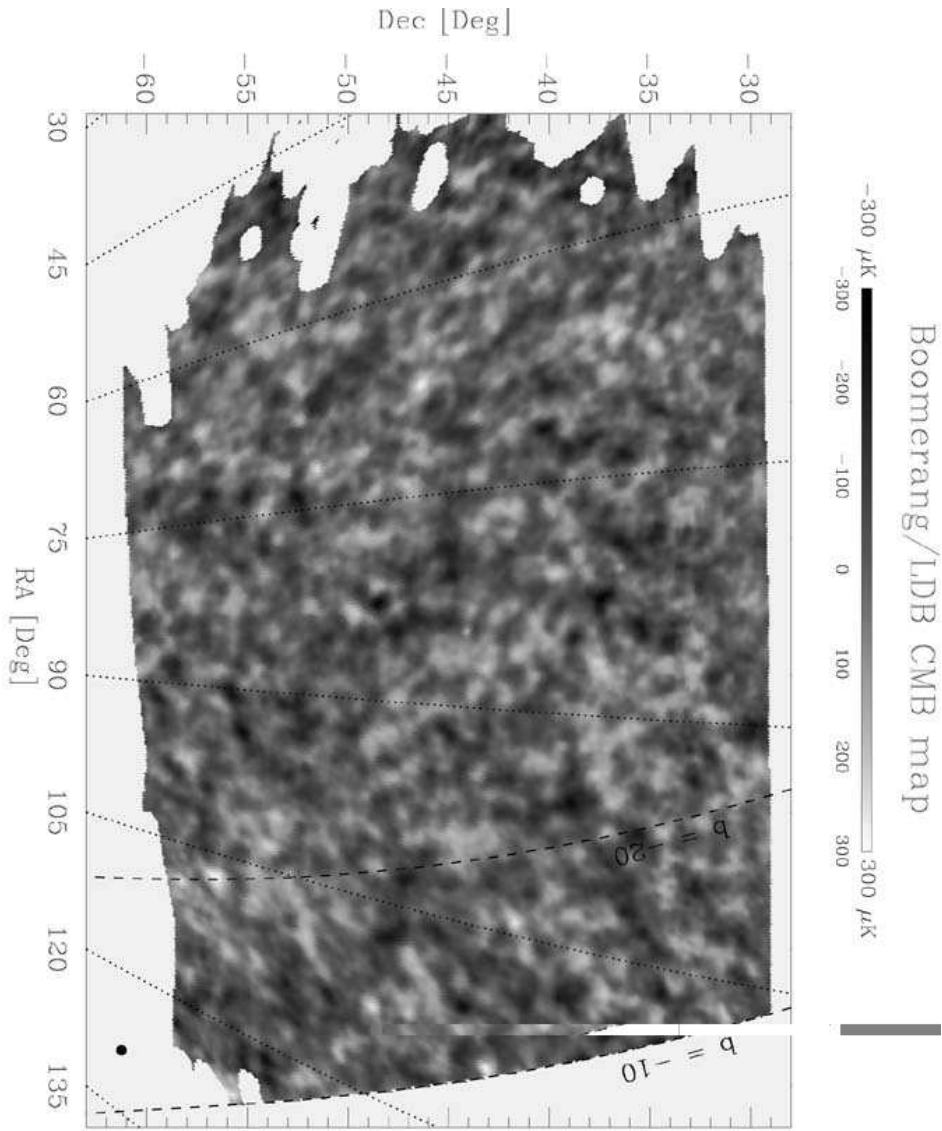}
\caption{
Map of $\sim 3\%$ of the sky at 150 GHz, as detected from three 
bolometers of  BOOMERanG. 
The data have been high-passed in the time domain. As
a consequence, structures larger than 10$^o$ have been efficiently
removed from the map. The map has HEALPIX 7' pixel size, and an
additional smoothing to 22.5' equivalent FWHM has been applied.
}
\end{figure}
We build maps from the scan data using both simple coadding in
pixels and an iterative method. The latter efficiently separates the noise
from the signal, producing an accurate estimate of the noise filter
(Prunet et al. 2000). This is then used in the
MADCAP maximum likelihood map estimator (Borrill 1999). Both methods use
the HEALPIX pixelization (Gorski et al. 1998), and find consistent results.
The map obtained coadding three of the 150 GHz channels is shown in fig.2.
The map features hundreds of structures with rms amplitude of $\sim 80 \mu K_{CMB}$
and typical size of the order of one degree. These are detected with high
S/N ratio. Consistency in amplitude and size with the maps obtained at 90 GHz and 240 GHz 
strongly suggests a cosmological origin of these strucutres. 
Three point sources in the map (highlighted by circles) 
are well known radio bright AGNs (flux $\sim 1 mJy$ $@ 230 GHz$) 
and have been used to check our pointing reconstruction. 
Our estimates of Galactic contamination is based on correlation
of our maps with the IRAS/DIRBE map (Schlegel et al. 1998) extrapolated 
to our frequencies, resampled along our scans and filtered in the time
domain in the same way as our channels. This exercise shows that 
in the observed region, at $b < -20^o$, Galactic
dust emission correlated with IRAS produces negligible signals at
150 GHz (Masi et al. 2000). The power spectrum of the central 
region shown in the map ($\sim 1\%$ of the sky) 
has been computed with two independent methods: a simple spherical harmonic
transform and the MADCAP maximum likelihood quadratic estimator (Borrill 1999).
They give consistent results. The power spectrum derived from one single 
150 GHz channel (the best one), using the 1$^o/s$ observations only,
is shown in fig.3. 
The power spectrum has been corrected for the transfer function of
the instrument, for the additional high-pass filtering, and for the
effect of the 14' HEALPIX pixelization. We performed several null tests to
check for systematic errors. The most important is the power spectrum
of the difference between the two maps obtained from the first half
of the 1$^o/s$ observations and from the second half. The power spectrum
is consistent with zero power, with a reduced $\chi^2 = 1.11$. Since the
payload moves by thousand km between the two halves of the data, 
any contamination from ground spillover should show-up in this 
difference spectrum, significantly increasing the $\chi^2$. Also
any contamination from the Sun in the far sidelobes of the telescope 
should produce similar effects, since the sun moves by many beamsizes 
during the measurements. The upper limit to systematic effects
obtained in this way is plotted in fig.3, where we also show
our estimates of contamination from Galactic dust and point sources.
We can safely conclude that the observed power spectrum is
not significantly contaminated by instrumental effects.
\begin{figure}
\plotone{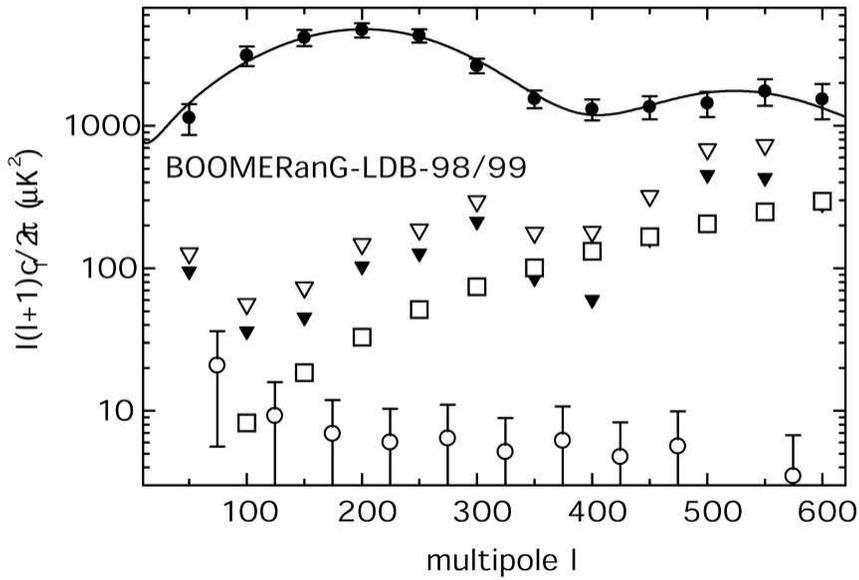}
\caption{
Power spectrum of 1$\%$ of the sky as measured at 150 GHz by a single
detector of the BOOMERanG experiment (filled circles). The continuous line
is the best fit in the framework of adiabatic inflationary models with cold 
dark matter and cosmological constant. The down triangles are the upper limits 
to systematic effects contaminating the data, estimated using jack-knife
techniques on data subsets (open= 95$\%$ c.l.; filled = 68$\%$ c.l.).
The open squares are our best estimate of power from point sources in the
observed field, based on WOMBAT (2000) extrapolation to 150 GHz of radio sources flux
in the PMN survey. 
The open circles are an estimate of the contribution from interstellar
dust emission correlated with the component mapped by IRAS. This has been
extrapolated from our 410 GHz dust monitor channel (Masi et al. 2000).
}
\end{figure}
\section{Cosmological Parameters}
This measurement of a well defined peak in the power spectrum of the CMB 
strongly suggests the presence of acoustic oscillations in the primeval
plasma (see e.g. Hu et al. 1997). In this scenario the degree-size
hot and cold spots evident in the maps are images of the acoustic horizons
on the last scattering surface at z $\sim$ 1100. Since CMB photons
traveled so long in the Universe, the average size of these
spots strongly depends on the average mass-energy density $\Omega$, which acts
as a magnifying ($\Omega > 1$) or demagnifying ($\Omega < 1$) lens.
Thus the location of the peak 
strongly depends on $\Omega$, i.e. on the curvature of the Universe, with
a flat geometry producing $\ell_{peak} \sim 200$.
If there is a non vanishing cosmological constant $\Lambda$, the 
relationship between $\ell_{peak}$, $\Omega$, and $\Lambda$ is non trivial
(Weinberg 2000; Hu et al. 2000). 
We have evaluated $\ell_{peak}$ by means of a quadratic fit to the
power spectrum data. We find $\ell_{peak} = (197 \pm 6)$. 
In the framework of inflationary adiabatic cold dark matter models this location
of the peak strongly suggests a flat geometry of the Universe 
(de Bernardis et al. 2000). More accurate statements require a 
through analysis of the full power spectrum dataset. A bayesian likelihood
analysis has been carried out in order to constrain instrumental and
cosmological parameters given the measured power spectrum, the
COBE power spectrum data at low multipoles, and a set of prior
distributions for the parameters. The result obtained depends on the
priors assumed. We carried out two independent analyses using different
parameters and various priors (Lange et al. 2000). In the first one we
use as cosmological parameters the physical density of baryons 
$\omega_b = \Omega_b h^2$ and of dark matter $\omega_c = \Omega_c h^2$;
the total mass-energy density $\Omega$, the energy density in 
cosmological constant $\Omega_\Lambda$, the spectral index of the spectrum
of primordial density perturbations $n_s$, the reionization optical depth 
$\tau_C$ and the overall normalization ${\cal C}_{10}$. In the second one
we used the parameters $h$, $\Omega_b$, $\Omega_c$, $\Omega_\Lambda$,
$n_s$ and ${\cal C}_{10}$. In both analyses we marginalize
over the uncertainty of the beam FWHM. We also marginalize over the
calibration uncertainty assuming a gaussian distribution. 
We compute a database
of power spectra for several millions of cases with different combinations
of the values of the parameters. The ranges and sampling 
selected for each parameter are wide enough to  
cover in detail the relevant parameters space.  
We then compute for each model the likelihood of the data, given 
the model and the assumed prior distributions. We use uniform prior 
distributions for all the assumed parameters. 
We use the offset lognormal approximation of Bond et al. (2000)
for the likelihoods. Finally, we compute
the likelihood distribution for the parameter of interest by
marginalizing over all the other parameters. 
The ranges of the parameters (equivalent to uniform priors) for analysis 1 are 
($\omega_c =$ 0.03, 0.8), ($\omega_b =$ 0.003125, 0.2),
($\Omega_\Lambda =$0, 1.1), ($\Omega_k = 1 - \Omega = $0.9, -0.5),
($n_s = $1.5, 0.5), ($\tau_c =$0, 0.5). For analysis 2 we have
($\Omega_m =$ 0, 1.1), ($\Omega_b =$ 0, 0.2),
($\Omega_\Lambda =$ 0, 0.975), ($h = $ 0.25, 0.95),
($n_s = $ 0.5, 1.5) . ${\cal C}_{10}$ is a continuous normalization
parameter in both cases. In fig.4 we plot the 
marginalized likelihood distribution for $\Omega$. In analysis
1 we only add "weak" priors for $h$ ($0.45 < h < 0.90$) and for the age
of the universe ($T > 10 Gy$). In analysis 2 use similar
priors: a gaussian prior for $h$ centered at 0.65 with dispersion 0.20, 
and the same $T > 10 Gy$ prior for the age. 
The effect of the applied priors has been analyzed in detail
in Lange et al. (2000). It is evident that a flat geometry ($\Omega = 1$) of the
Universe is consistent with the BOOMERanG power spectrum of fig.3. 
The comparison of the two marginalized likelihoods shows the 
robustness of the result against changes in the details of the 
analysis method. 
\begin{figure}
\plotone{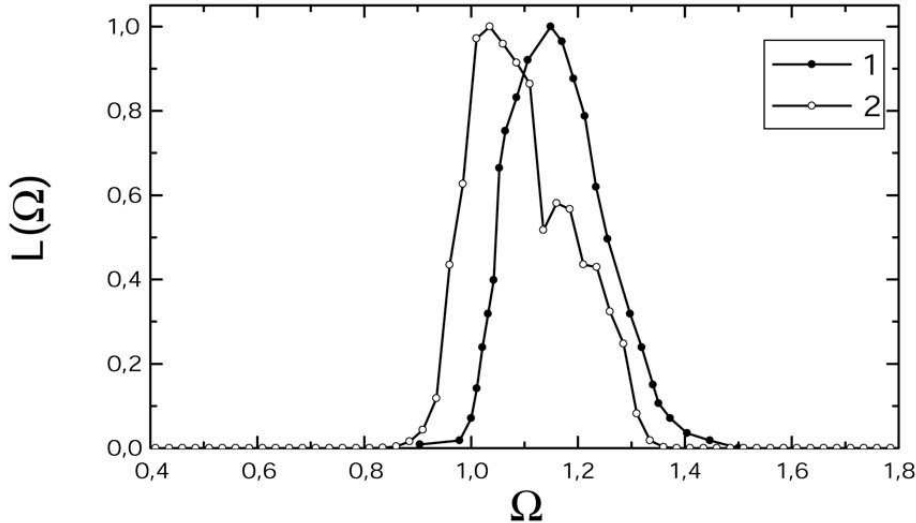}
\caption{
Likelihood of $\Omega$ from BOOMERanG and COBE. Curves 1 and 2
differ due to the different priors applied and parameters used 
(see text).
}
\end{figure}
There is an important parameter degeneracy that contributes to the width
of the likelihood distribution for $\Omega$ (Bond and Efstathiou 1999).  
For example, given a flat model with $\Omega_\Lambda \sim 0.3$ 
other models (both open and closed) with virtually identical CMB power 
spectra can be found by fixing $\Omega_b h^2$ and $\Omega_c h^2$, while 
changing $\Omega_\Lambda$ and $h$ in a
coordinated way.  Our analysis is Bayesian, so the details of the
likelihood curve will then depend on the density of models
found in each direction along that particular path. In both analyses
we find a number of closed models larger than the number of 
open models, and the shift of the two curves is due to the density
of models considered in the two cases.
The BOOMERanG data also constrain other cosmological parameters. In this
framework the density of baryons controls the relative amplitude of the
second acoustic peak of the power spectrum with respect
to the amplitude of the first peak. Compressions of the plasma are favoured 
with respect to rarefactions when the density of baryons is increased.
As a consequence, the second peak is suppressed relative to the first one.
In the BOOMERanG spectrum discussed here the second peak is not evident,
and we can only set an upper limit (an a lower one as well) to its amplitude.
Models with a density of baryons higher than 
the Big Bang Nucleosynthesis (BBN) value $\Omega_b h^2 = 0.019 \pm 0.002$ 
(see e.g. Tytler et al. 2000) fit the data better.  
Using "weak" priors we get $\Omega_b h^2 = ( 0.036 \pm 0.006 )$, a 
value $\sim$ 2.8$\sigma$ higher than the BBN value. Assuming
$\Omega = 1$ and weak priors on $h$ and $T$ we get $\Omega_b h^2 = ( 0.031\pm 0.004)$,
still $\sim 3\sigma$ off from the nucleosynthesis one.
We must stress the fact that we are comparing two very different measurements of the
baryon density: the CMB one is obtained from the density of
baryons at $z \sim 1000$, while the BBN one is inferred from measurement
of the abundance of Deuterium in the line of sight towards three
clusters at $z \sim 3$, and is representative of the primordial abundances.
In the first case the physical processes involved are acoustic oscillations
in the plasma before recombination, while in the second case are nuclear 
reactions occuring a few minutes after the big bang. 
It is rather spectacular that measurements
obained with methods which are so orthogonal and subject to
completely different sytematics produce results so close to
each other. 
Another parameter constrained by the BOOMERanG power spectrum
is $n_s$. Using weak priors we get $n_s = (1.04 \pm 0.09)$, in good
agreement with the simplest inflationary scenario. 
A way to move the CMB measurement of the baryons to be more consistent 
with the BBN is to assume a shallower initial spectrum ($n_s \sim 0.9$):
this is a simple way to reduce the height of the second peak
without needing too many baryons. 0.9 is not the best
"marginalized" estimate for $n_s$, but the fit to the data is 
very good anyway (see e.g. Lange et al. 2000; see also Tegmark et al. 2000
where large scale galaxy distribution data and CMB data are fitted simultaneously).
A value $n_s \sim 0.9$ is also consistent with modern inflationary models
(see e.g. Kinney et al. 2000).
The parameters $\Omega_\Lambda$ and $\Omega_m = \Omega_c + \Omega_b$ 
are strongly degenerate: combinations of these parameters resulting in
the same $\Omega$ produce very similar power spectra (Bond and Efstathiou 1998).
This "geometrical degeneracy" is broken once priors deriving from
independent observations are used. We find that using either
a prior due to observations of distant supernovae (Riess et al. 1998; 
Perlmutter et al. 1999) or a prior due to observations of the large scale
distribution of galaxies (see Lange et al. 2000) we get several $\sigma$ detections of
$\Omega_\Lambda = (0.66 \pm 0.07) $ and  $\Omega_m = (0.48 \pm 0.13)$.
\section{Conclusions}
The data from BOOMERanG strongly support two of the main predictions of
the simplest inflationary models: a nearly flat geometry of the Universe 
and a nearly scale-invariant spectrum of primordial density fluctuations. 
The detected density of baryons is from higher to consistent with 
that predicted from BBN, depending on the assumptions for other parameters.
Combined with observations of different nature the BOOMERanG data 
produce significant detections of 
$\Omega_\Lambda$ and $\Omega_m$. These conclusions
are all confirmed by a joint analysis of the COBE, BOOMERanG and
MAXIMA-1 data (Jaffe et al. 2000). Further insights are expected
from the analysis of the data from the other detectors in BOOMERanG, 
from the data of the second flight of MAXIMA, and from the data
of several interferometric CMB experiments underway.
\section{Acknowledgments}
The BOOMERanG project has been supported by PNRA, Universit\'a ``La Sapienza'', 
and ASI in Italy, by NSF and NASA in the USA, and by PPARC in the 
UK. We would like to thank the entire staff of the NSBF, and the US 
Antarctic Program personnel in McMurdo for their excellent 
preflight support and a marvelous LDB flight. DoE/NERSC 
provided the supercomputing facilities. 
Web sites: (http://oberon.roma1.infn.it/boomerang) and 
(http://www.physics.ucsb.edu/$\sim$boomerang).

\end{document}